# Using Ethnographic Methods to Classify the Human Experience in Medicine: A Case Study of the Presence Ontology

**Amrapali Maitra, M.D., Ph.D.[1,2], Maulik R. Kamdar, Ph.D.[3], Donna M. Zulman, M.D., M.S.[4,5], Marie C. Haverfield, Ph.D.[6], Cati Brown-Johnson, Ph.D.[4], Rachel Schwartz, Ph.D.[7], Sonoo Thadaney Israni, M.B.A[2], Abraham Verghese, M.D.[2], Mark A. Musen, M.D., Ph.D.[3]**

[1] Department of Medicine, Brigham and Women's Hospital, Boston, MA, USA
[2] Stanford Presence Center, School of Medicine, Stanford University, Stanford, CA, USA
[3] Center for Biomedical Informatics Research, Stanford University, Stanford, CA, USA
[4] Division of Primary Care and Population Health, Stanford University, Stanford, CA, USA
[5] Center for Innovation to Implementation, VA Palo Alto Health Care System, Menlo Park, CA, USA
[6] Department of Communication Studies, San Jose State University, San Jose, CA, USA
[7] WellMD Center, Stanford University School of Medicine, Stanford, CA, USA

## Abstract

**Objective**
Although social and environmental factors are central to provider-patient interactions, the data that reflect these factors can be incomplete, vague, and subjective. We sought to create a conceptual framework to describe and classify data about *presence*, the domain of interpersonal connection in medicine.

**Methods**
Our top-down approach for ontology development based on the concept of "relationality" included the following: 1) broad survey of social sciences literature and systematic literature review of >20,000 articles around interpersonal connection in medicine, 3) relational ethnography of clinical encounters (*n*=5 pilot, 27 full) and 4) interviews about relational work with 40 medical and nonmedical professionals. We formalized the model using the Web Ontology Language in the Protégé ontology editor. We iteratively evaluated and refined the Presence Ontology through manual expert review and automated annotation of literature.

**Results and Discussion**
The Presence Ontology facilitates the naming and classification of concepts that would otherwise be vague. Our model categorizes contributors to healthcare encounters and factors such as Communication, Emotions, Tools, and Environment. Ontology evaluation indicated that Cognitive Models (both patients' explanatory models and providers' caregiving approaches) influenced encounters and were subsequently incorporated. We show how ethnographic methods based in relationality can aid the representation of experiential concepts (e.g., empathy, trust). Our ontology could support informatics applications to improve healthcare such annotation of videotaped encounters, clinical instruments to measure *presence*, or EHR-based reminders for providers.

**Conclusion**
The Presence Ontology provides a model for using ethnographic approaches to classify interpersonal data.

# 1 INTRODUCTION

Modern medicine has advanced the treatment of disease but at times infringes on the simple ritual of doctors using compassion, listening, and skilled touch in the bedside exam to connect with patients[1,2]. The emotional labor of creating connection is an important part of a healthcare provider's role, yet such care is "far more complex, uncertain, and unbounded than professional medical and nursing models suggest"[3]. Investing in interpersonal connection may prevent burnout of healthcare providers and increases patient satisfaction[4,5], yet it is challenging in our technologized era of medicine[2,6,7].

*Presence* is an emerging medical discourse that refers to the "purposeful practice of awareness, focus, and attention with the intent to understand and connect with patients"[8]. There is currently no unifying framework to describe, capture, and classify human and environmental data surrounding interpersonal connections in clinical encounters, and the interpersonal interactions that comprise *presence* cannot be gleaned from the EHR[9]. We define "interpersonal" as "a selective, systemic process that allows people to reflect and build personal knowledge of one another and create shared meanings"[10]. The *presence* domain is of increasing relevance to informatics research and applications that seek to improve the individual experience of healthcare and delivery systems[11] through electronic health record (EHR) innovations, scribe programs for documentation, or integration of smartphones into clinical care.

Data related to human experiences and social interactions are often incomplete and sometimes subjective; they are documented qualitatively in multiple, idiosyncratic, and partial ways. Such data arise through interactions among different individuals, with diverse objects, across multiple physical and virtual spaces. An example is the Social History within the EHR where demographic data like marital/partner status, occupation, substance use, and sexual history are listed in series of drop-down boxes rather than elaborated as an opportunity to situate medical complaints within patients' complex life circumstances[12].

In this study, we combine biomedical ontology engineering with ethnographic methods to define the factors contributing to interpersonal connection in medicine. Specifically, we have conceptualized and developed the Presence Ontology, a systematized vocabulary of terms that models the interactions taking place every day among healthcare providers, patients, and their families and friends. Developing a conceptual vocabulary for *presence* could generate informatics innovations to better evaluate the patient experience including satisfaction[13]; mitigate clinician burnout and support joy of practice[14,15]; and equitably deliver personalized care in the artificial intelligence (AI) transformation of medicine[16].

The ontology was developed through interdisciplinary collaboration of experts in medicine, bioinformatics, anthropology, linguistics, communication, psychiatry, and public health. Our approach may provide clarity and consensus to the important but ill-defined domain of human experience in bioinformatics and has relevance to informatics subfields where interpersonal data are central to knowledge classification domains.

# 2 CLASSIFYING PRESENCE

We sought to identify the elements of interpersonal connection in the patient-physician relationship and engineer them into an explicit formal specification called an ontology. By utilizing a shared language with defined relationships, we strove to make subjective data and metadata in healthcare more expressive and precise—such data are often taken to be a black box by computational researchers because they are subjective and may be vaguely defined. The development and use of ontologies for clinical care is a critical requirement in the creation of automated decision support tools and clinical research databases for data harmonization and semantic interoperability[17]. An example of ontology engineering of broad clinical concepts is the widely used and exhaustive vocabulary known as the Systematized Nomenclature of Medicine – Clinical Terms (SNOMED CT)[18].

In order to model human experience in clinical encounters, we combine the conceptual abstraction of social sciences theory with the granularity of ethnography. Literature from sociology, anthropology, and linguistics provides fertile conceptual terrain to describe human experiences related to healthcare. Ethnographic methods are of increasing interest in bioinformatics[19,20]; they are fine-grained and describe the variability of experience. Our methods build on ethnographic approaches for developing electronic knowledge bases[21–23].

Our ontology shares some categories with existing frameworks[24–26] including *characteristics* (identity features that define both patients and providers), *encounters* (instances where multiple people come together) and *emotions* (intrapersonal experiences through which interpersonal experience is mediated), but brings additional rigor by naming terms with logic and consistency for usability across many providers and interactions. We extend Ventres and Frankel's "shared presence" framework focused on providers' behaviors and actions (e.g. to listen, examine, educate) by incorporating the behaviors and qualities of patients and influence of environments in shaping *presence* within clinical encounters[24]. We also build on Larson and Yao's model of empathy which describes how antecedents (e.g. physician or patient characteristics or situational characteristics) affect empathic processes, which in turn result in intrapersonal and interpersonal outcomes that extend to physician and patient outcomes (e.g. burnout, patient satisfaction, and healthcare outcomes)[25]. While the framework models clinical encounters as linear and one-dimensional, we elaborate further to account for the multiple, intersecting ways in which people's characteristics, environments, and behaviors coalesce to shape *presence*.

The ontology leverages the Presence 5 framework, which describes evidence-based practices that promote clinician presence: 1) prepare with intention, 2) listen intently and completely, 3) agree on what matters most, 4) connect with the patient's story, and 5) explore emotional cues.[9] These recommendations embed concrete, measurable actions and behaviors within the clinical encounter, such as use of time, body position, management of a computer screen, and communication style. Formative research for the Presence 5 framework included a systematic literature review, observations of clinical encounters, and interviews with physicians and non-medical professionals, data, which also informed ontology development **(Section 3**).

# 3 METHODS

We applied ethnographic principles to ontology engineering centered on the concept of relationality. We identified and modeled key concepts and relations surrounding the domain of *presence* and developed the Presence Ontology into a formal, usable clinical artifact. We followed a top-down approach for ontology development starting with the highest-level (most abstract) concepts in our domain and then defining sub-concepts. Domain analysis was based upon literature survey, ethnographic observations of clinical encounters, qualitative insights from professionals engaged in relational care, and meetings of clinical and research experts over nine months.

## 3.1 Broad survey and systematic review of domain literature

We identified preliminary concepts pertaining to the domain of *presence* through a broad survey of literature in both medicine (using *PubMed*[27]) and the social sciences (journals and books of anthropology, communication, sociology, and psychology) around topics of interpersonal connection. Keywords for the broad literature survey included Patient-Physician Relationship, Communication, Empathy, Power Dynamics, Patient-Centered Care, Technology, Burnout, Trust, and Mindfulness. The broad survey allowed for attention to concepts that may not readily appear in the conventional medical literature, such as the cultural and political milieu of healthcare, the spoken and unspoken aspects of clinical care, as well as the hierarchies, language, and feelings that structure encounters. Key concepts of relevance to *presence* included the clinical encounter as ritual[1,28], power dynamics in medicine[29–31], the emotional labor and ethical stance of care work[26,32], the illness experience and explanatory models[31,33], and the impact of technology and the built environment on the connection between patients and providers[34,35]

The broad survey was expanded through a systematic literature review conducted for the Presence 5 study[9,36]. Three databases were searched across biomedical and social sciences (*PubMed*[27], *EMBASE*[37], and *PsycINFO*[38]) capturing research from January 1997 to August 2017 for randomized controlled trials and controlled observational studies of evidence-based interpersonal interventions geared toward improving presence that included at least one outcome measure of the "quadruple aim" (i.e., patient health outcomes, patient experience, clinician experience, or cost)[36,39]. A broad array of MeSH terms and keywords encompassing domains such as trust, empathy, humanism, and communication were used. The review yielded 21,835 articles; 77 of which were retained after screening of titles, abstracts, and full texts (and 73 of which are the focus of a published systematic review[36]). For ontology development, the 77 papers were reviewed in-depth for content and conceptual language related to *presence*. The abstracts were also used in a later stage of ontology evaluation and revision (See **Section 4.5**).

The review began with 21,835 articles. After screening of titles, abstracts, and full texts (including systematic reviews for component studies), 77 unique studies were preliminarily identified for quality assessment and data extraction stage (in the published systematic review, 73 of the 77 articles were retained). These papers were reviewed for content and conceptual language related to *presence*. The abstracts were also used in a later stage of ontology evaluation and revision (See **Section 4.5**).

## 3.2 Relational ethnography to develop upper-level categories and concepts

The literature survey resulted in an initial conceptual sketch of the domain, modeled around three key components of clinical encounters: *1)* providers, *2)* patients, and *3)* environments in which encounters take place. However, subsequent discussion exposed many gaps; the model did not allow us to speak about factors generated *through* interaction (such as trust) as well as identities shared *across* types of persons (both providers and patients, for example). Seeking new ideas for upper-level abstraction, we used ethnographic methods to elaborate the configuration of *presence* encounters.

Relational ethnography of clinical encounters ($n$=5 pilot, 27 full) allowed us to examine tension, incompleteness, or unexpectedness. Relationality provided a conceptual foundation to situate patient encounters as hierarchical, interpersonal, spontaneous, and unbounded, rather than modeling patients as fixed entities with rigid roles. We use "relationality" as defined by sociologist Desmond, as a focus on understanding processes, fields, and conflicts[40]. Relational ethnography conceptualizes interactions beyond the boundaries of place or group; instead, it seeks to "broaden and expand" the field of objects and relationships therein. This method is useful for health informatics because it is grounded in practice theory, attending to the ways in which "technology and social practice are mutually elaborated" in clinical interactions[41].

An anthropologist observed five pilot clinical encounters with IRB approval (Stanford Institutional Review Board Protocol No. 30711). Written informed consent was obtained from patients prior to observations. Observations were centered on the elements of *presence* in patient-physician interactions with one provider at a medicine subspecialty clinic in California. Fieldnotes focused on structure and content of conversation, body position, movement and touch during the physical examination, and patterns of speech and silence. Patients were interviewed post-visit using a semi-structured interview script to elicit perceptions about the level of connection to provider, feelings of vulnerability, perceptions of care and empathy, and satisfaction.

To validate findings from the pilot observations, we analyzed data from 27 patient-physician encounters that were conducted for the Presence 5 study described previously[9]. These encounters spanned three primary care settings in California (an academic medical center, a Veterans Affair facility, and a federally qualified health center) and used a "rapid ethnography" approach[42] centered around a conceptual model for Presence observations **(Figure 1**). Methods included observation and video- or audio-recording of clinical encounters, fieldnotes, team debriefing, post-visit interviews with clinicians and patients about strategies to foster *presence*, and consensus coding[9]. Procedures were approved by Stanford University IRB (Protocol No. 42397). Physicians and patients provided written informed consent for observations, recordings, and interviews.

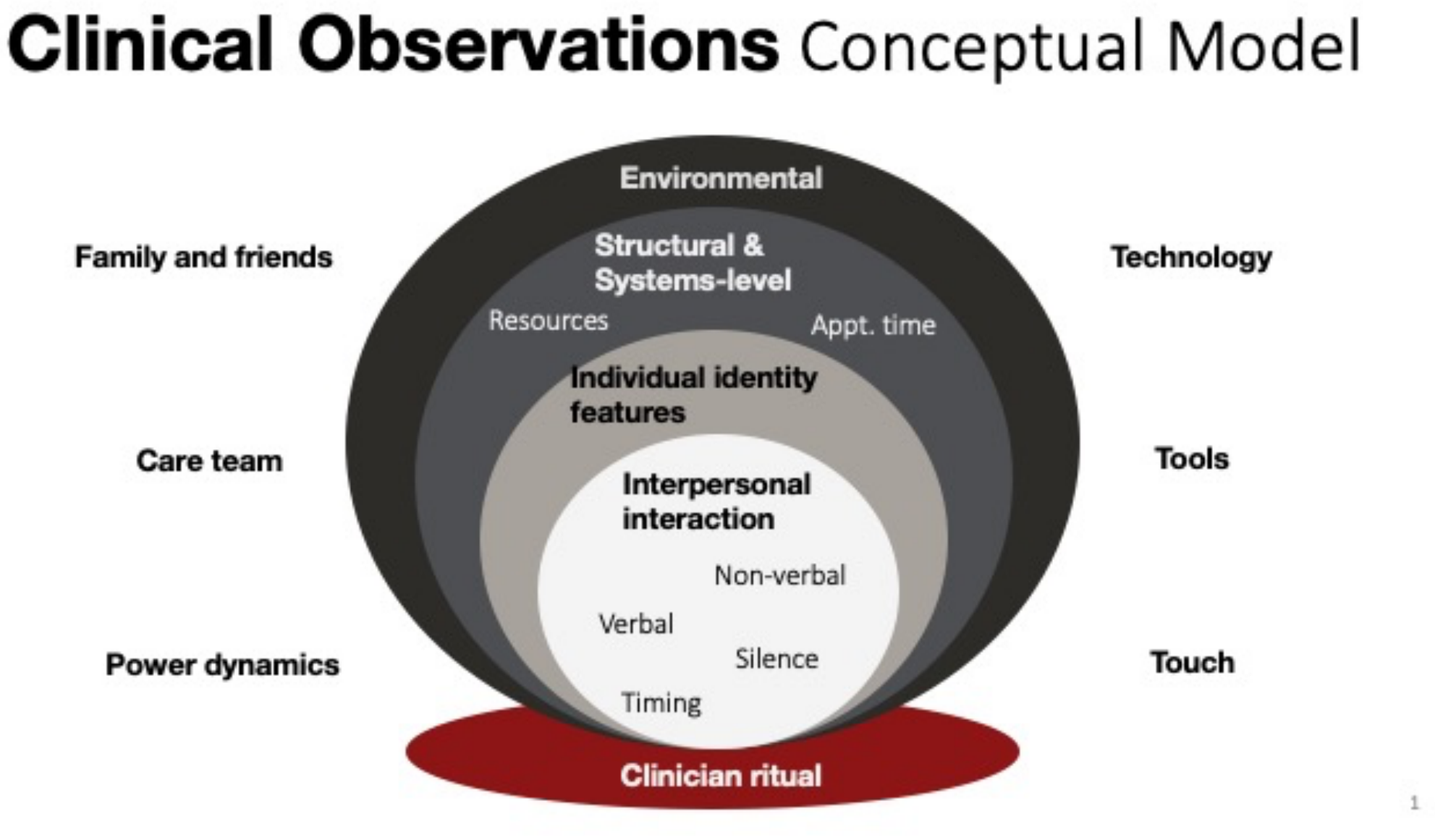

**Figure 1. Conceptual model for clinical observations to develop *Presence* relational ethnography.** The model depicts the conceptual hierarchy for themes related to the clinical encounter. The model was developed through literature survey and expert review and finalized prior to conducting the relational ethnography. Presence research team members were trained using this model in order to structure fieldnotes for observed patient-physician encounters (*n*=27) using a rapid ethnography approach. At the core is the clinical ritual, upon which is layered interpersonal interaction (with attention to verbal and non-verbal communication, timing, and silence), then individual identity features of both the clinician and patient, structural and systems-level features such as clinic resources or wait time, and finally the environmental milieu within which encounters occur. Additional elements that mediate the encounter include power dynamics, care team members, the patient's family and friends, technology, tools, and touch.

## 3.3 Qualitative insights from relational care professions to refine categories

The ethnographic study was supplemented by trans-disciplinary qualitative insights, leveraging data from formative research for the Presence 5 study, in which purposive sampling was used to identify 30 professionals from outside the field of medicine whose work involves relational care, such as police officers, personal care services (e.g. yoga, massage therapy), management fields (CEO, school principal), education, the arts, and social services[9,43]. Participants were interviewed about their approaches to interpersonally intense encounters. An inductive analysis was used to identify themes. 10 physicians were also interviewed to compare and correlate themes[8]. Themes included conscientious approaches to self-care that permit greater *presence*, protected time for peer interaction, and an emphasis on fostering a bidirectional exchange to enhance professional fulfillment. These insights refined our upper-level categories of *presence*.

## 3.4 Collective domain expertise of experienced clinicians and researchers

Collective expertise was elicited from a team of researchers whose backgrounds encompass medicine, bioinformatics, anthropology, communication, linguistics, psychiatry, and public health. The research team met biweekly over six months to iterate on conceptual categories and relationships within the ontology. At each stage, the team of experts reviewed evidence to date, refined subsequent methods, and discussed applicability of concepts (through reflection on moments of *presence* breakdown in individual clinical experience, for example). This allowed for synthesis of mixed methods to generate ontology concepts.

## 3.5 Formalization of ontology using OWL in Protégé

The ontology was implemented using OWL (Web Ontology Language) in Protégé, one of the most widely used open source ontology editors[44]. Using the preliminary steps, we enumerated different class entities and properties. We modeled these entities using appropriate OWL axioms such as classes, object properties, data properties, and annotations. Then, following best practices[45], we externally cross-referenced relevant terms from two existing ontologies, SNOMED CT and the Emotion Ontology[18,46]. We did not import either ontology because the contexts in which they were developed are different from our own—that is, SNOMED CT is supposed to be an exhaustive clinical reference terminology, whereas the Emotion Ontology is focused heavily on emotions rather than clinical encounters or their participants. While we could have extracted modules from SNOMED CT for import within the Presence Ontology, such modules could have led to semantic inconsistency. Finally, we uploaded the Presence Ontology in the BioPortal repository[47], the world's largest open source repository of biomedical ontologies, for annotations and dissemination.

## 3.6 Evaluation and revision of ontology from the *presence* systematic review

A corpus of abstracts from the systematic review of *presence* (**Section 3.1**) was used to evaluate the Presence Ontology[36]. We conducted a manual formative assessment of the abstracts along with a review from clinician experts to identify key conceptual gaps in our initial ontology. We also completed a data-driven summative evaluation using the Presence ontology (uploaded in the BioPortal repository) and the BioPortal Annotator[48], which is widely used for the annotation of biomedical texts and electronic health records with UMLS concepts. The BioPortal Annotator takes as input a dictionary of term labels as well as a set of their synonyms (e.g., "Clinician," "Physician," and "Doctor") and generates annotated abstracts as output. We annotated 77 abstracts with 306 terms (classes and object properties) from the Presence Ontology. We also identified terms that frequently co-occur (i.e., mentioned together in the same abstract) with a given term.

# 4 RESULTS

## 4.1 Presence Ontology

Our ontology models *presence* within clinical medicine. Patients and providers are involved in encounters, which contain several subparts. Encounters are influenced by various personal, environmental, and relationship factors, as well as emotions, qualities, and characteristics of the participants. Encounters often result in health outcomes. They also produce qualities that set the stage for future encounters. Interactions are also modeled as time-bound (with a start time and an end time).

A diagrammatic representation of the model behind the Presence Ontology is illustrated in **Figure 2.** The colored regions in the representation represent the different upper-level class entities with the subclasses listed in those boxes. Object properties are italicized alongside the arrows between class entities.

Our model includes the following upper-level class entities shown in the colored boxes:

- **Person** - All individuals and specific Healthcare Roles like Patients, Providers, or "Framily" (i.e., Friends and Family) involved in an encounter
- **Characteristic** - Defining features of individuals such as age, gender, race, etc.
- **Encounter** - A subclass of Events, which involves at least one Provider and at least one Patient. Encounters can consist of multiple sub-encounters and smaller interactions or "Encounter Components". Encounters result in outcomes for patients, providers, and healthcare systems.
- **Action** - Each Encounter Component may involve several actions (e.g., patient workup) that are performed by the individuals involved in the Encounter
- **Object** - An Action often involves the use of certain tools, and on the opposite spectrum, an Action may be interrupted by certain Objects (e.g., interruption by pager alert)

- **Factor** - Encounters between a patient and a provider may be influenced by several external factors, which often include Communication (Verbal, Nonverbal, Paralinguistic), the nature of the Patient Provider Relationship, elements of the Patient History, and features of the Environment.
- **Quality** - Non-relational qualities (e.g., experienced internally by a single individual) and relational qualities (e.g., interpersonal connection experienced by two or more individuals) can be generated through Encounters or can influence them.
- **Emotion** - Encounters are also influenced by emotions of the different participants in that encounter.

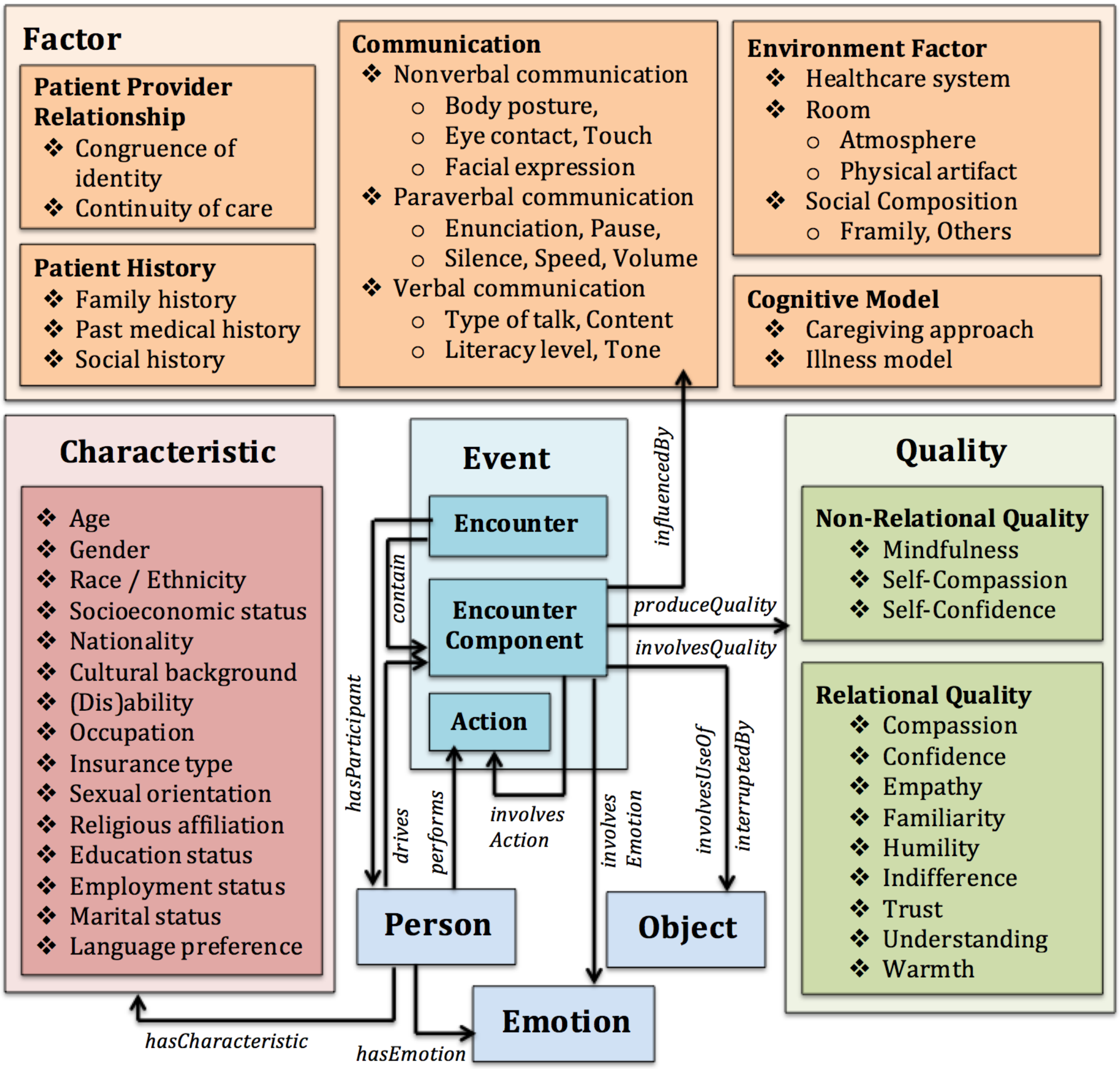

**Figure 2. Class diagram of the Presence Ontology elaborating upper-level class hierarchies and object properties.** Each colored box indicates an upper-level class (e.g., "Factor"), with subsequent inner boxes depicting the hierarchy under the upper-level class (e.g., "Patient History" is a subclass of "Factor", and in turn "Family History" is a subclass of "Patient History"). Different classes are connected using object properties in the Presence Ontology. For example, the object property "performs" associates the class "Person" with the class "Action" (an individual under the class "Person", whether a "Patient" or a "Provider", will perform some "Action"), whereas, the object property "hasCharacteristic" associates the class "Person" with the class "Characteristic" (an individual under the class "Person" has at least one "Characteristic", such as "Age", "Occupation", "Race", etc.). For simplicity, we have only shown the most relevant classes in this class diagram and refer the reader to explore the Presence Ontology in the BioPortal repository for more information around the class hierarchies and the object properties.

The Presence Ontology allows for naming and classification of concepts that would otherwise be vague. The ontology also allows mapping of the relationships among these disparate components in an encounter. Our ontology uses the object property "produceQuality" to model how a relational quality like compassion is *produced* within a specific interaction. This is depicted in **Figure 2** where the "produceQuality" object property is associated with the classes "Encounter Component" and "Quality", where "Quality" is further subcategorized into relational (e.g., compassion) and non-relational qualities (e.g., self-confidence). Recent ethnographic studies of medical encounters support the idea that qualities like compassion are both "dispositional" (an individual quality or characteristic) and "situational" (related to the encounter context)[49].

The Presence Ontology also suggests new conceptualizations of technology and time in clinical encounters. The ontology uses object properties such as "InvolvesUseOf" and "InterruptedBy" to describe negative and positive ways a technological object (computer, smartphone, etc.) could act within an encounter. Thus, it offers a way to integrate data around technologies or devices in healthcare as both a tool and a barrier to developing interpersonal connection. Second, conceptualizing medical encounters as time-bound entities allows clinical applications like tracking time-related data. In medicine, encounters length is an important variable, as time (e.g., length of encounter, wait time) may affects patients' perceptions of the interaction. In one study of 5,000 patients concerning prior healthcare encounters, time with physicians was a stronger predictor of patient satisfaction than wait time[50].

## 4.2 Ontology Evaluation and Revision

### 4.2.1 Formative Evaluation

Through the manual formative assessment of the initial ontology and the abstracts extracted from the systematic review, we identified a key conceptual gap in our ontology: cognitive models. We revised the ontology to include Cognitive Model as a subclass of Factors that influence Encounters. Cognitive Model comprises both Caregiving Approaches (approaches held by providers such as patient-centered care, shared decision-making, or motivational interviewing) and Illness Models (also termed illness representations or illness scripts) conceptualized by patients. While intrapersonal experiences like emotions had been modeled, cognitive processes were an important addition as they indicated how attitudes regarding care or illness may shape encounters.

### 4.2.2 Summative Evaluation

**Figure 3** depicts the top 20 terms from the Presence Ontology that frequently appeared in our corpus of 77 abstracts. Each red bar indicates the total number of abstracts in which the represented term is mentioned, whereas the corresponding blue bars indicate the total number of other terms from the Presence Ontology with which the represented term co-occurs in the corpus (i.e. in a given abstract, the represented term is identified with a few other terms from the Presence Ontology). For example, from the histogram, it is apparent that the concept of "Patient" is identified in more than 70 abstracts and co-occurs with approximately 90 other terms from the Presence Ontology.

The systematic review identified papers with a high patient-centric focus, which previous models in the domain of *presence* often lack. Through this histogram, it is evident that the Presence Ontology is able to identify concepts such as "patient-centered care", "patient satisfaction", "confidence", and "stress," which focus more on the qualities and emotions that are the outcome of patient-provider encounters. The inclusion of these patient-centric terms in the Presence Ontology broaden the focus from providers as the only possible driver of interactions, a focus that is prevalent in existing models, toward a relational view of encounters. Finally, concepts which refer to "Cognitive Models" ($11^{th}$ most common term in **Figure 3**) and were identified in the formative assessment are often mentioned in literature surrounding presence.

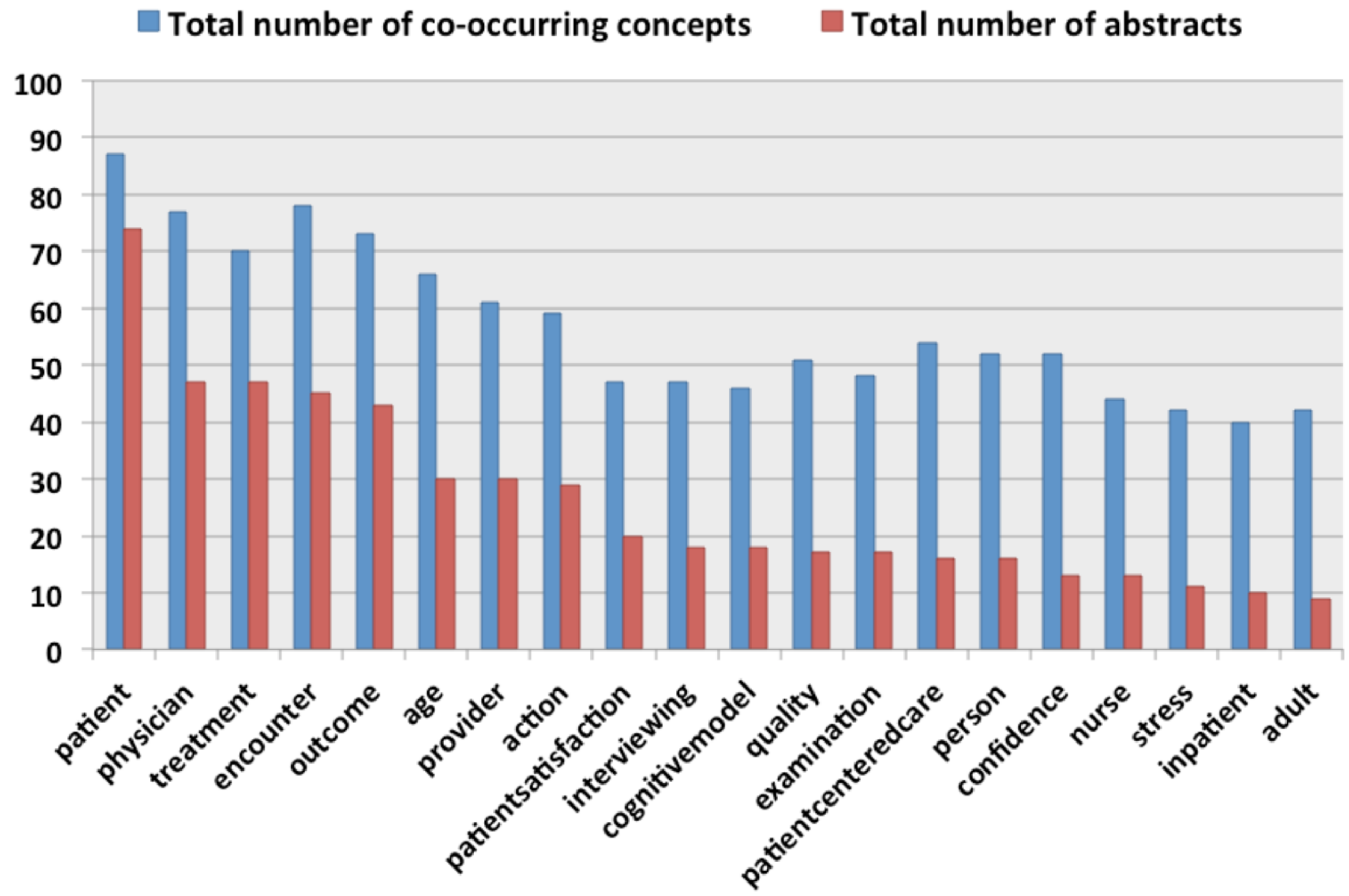

**Figure 3. Concepts from the Presence Ontology that were most commonly identified in the 77 Abstracts related to *Presence* literature.** The X-axis in this histogram showcases the 20 most commonly identified concepts from the Presence Ontology in the *Presence* literature. Each red bar in the histogram indicates the total number of abstracts in which the represented concept is mentioned, whereas each blue bar indicates the total number of Presence concepts that co-occur with the represented concept in these abstracts. The importance of patient-centric concepts in our approach broadens the focus from providers as drivers of human experience in medicine toward a relational framework for *presence*, the direct outcome of our ethnographic methods for ontology development.

Using this evaluation approach and the Presence Ontology, we can construct *presence* co-occurrence networks to understand which factors and qualities are simultaneously experienced in clinical encounters. For example, as shown in **Figure 4**, the emotion "Stress" is mentioned in 13 abstracts, and in those abstracts it co-occurs with approximately 45 other terms from the ontology. "Stress" frequently co-occurs with both "Patient" and "Physician," as well as with the concepts of "Outcome," "Confidence," "Anger," "Empathy," and "Trust". The co-occurrence analysis can provide insight into features that may diminish *presence* in encounters.

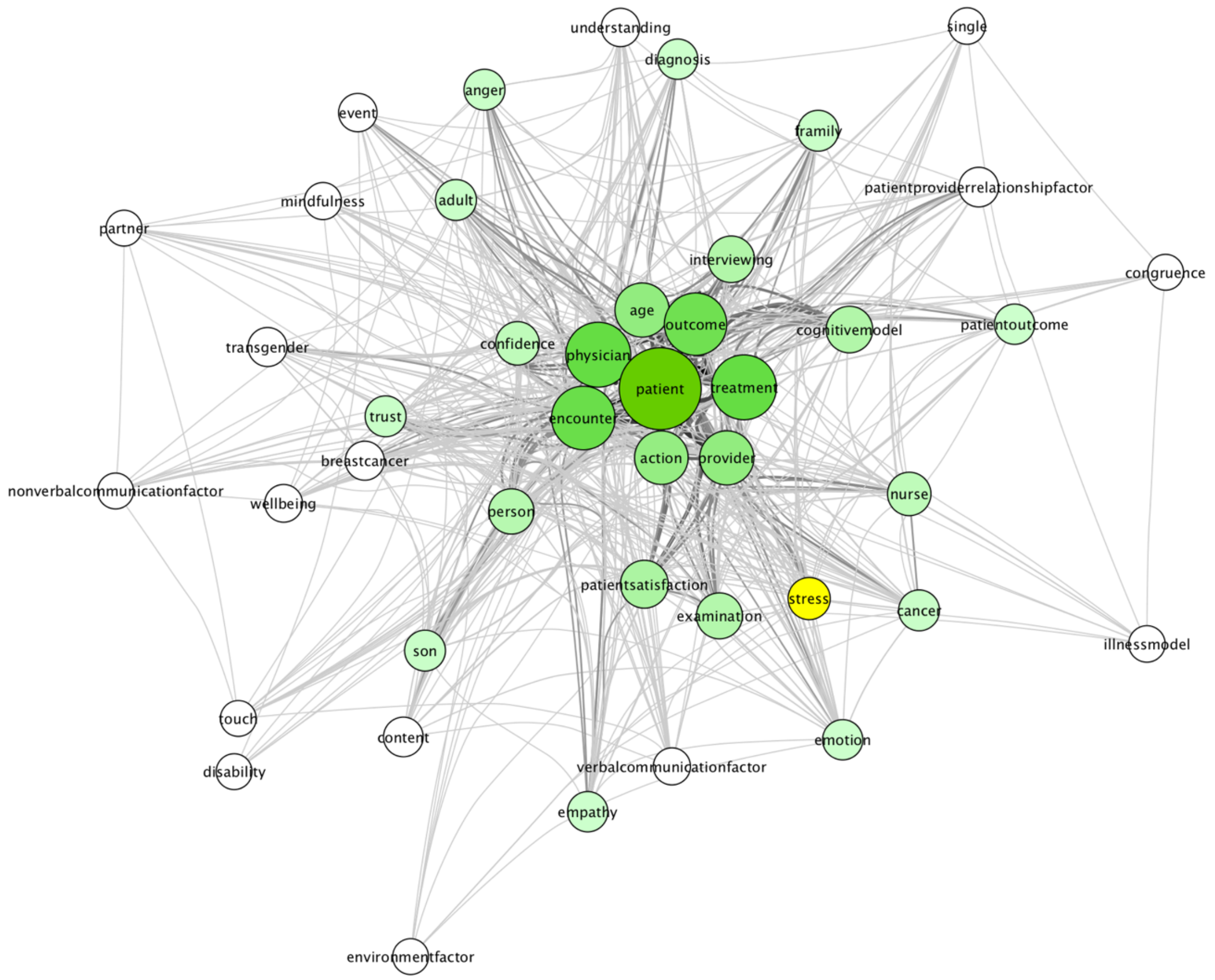

**Figure 4: Co-occurrence network for the concept of "stress" generated from the identification of Presence concepts in literature.** The concept of "stress" from the Presence Ontology generally co-occurs with common concepts of "patient" or "physician", but also with concepts such as "empathy", "anger", "trust", etc. The size and the color of the nodes is indicative of the number of abstracts in which the *presence* concept is identified, and the thickness of the connecting edges between two nodes is indicative of the number of abstracts in which the connected concepts co-occur together.

# 5 DISCUSSION

The specificity of controlled vocabularies can push forward the fields of medicine and clinical informatics. We have sought to introduce specificity to the domain of *presence* with challenging data qualities that are subjective (based on individual perception and experience), partial (not all information about the domain can be known), unpredictable (both patients and providers often improvise in interactions), and ever-changing (any changes in social world produces a change in the domain of human connection).

The ethnographic approach of relationality, grounded in practice theory, provides a useful model for development of knowledge systems that strive for ontological realism while remaining rooted in the core of healthcare: human interactions occurring over time. The role of realism in ontologies has been a debate in clinical and biomedical informatics; Smith and Ceusters argue that ontologies should comprise universals taken from an objective reality[51]. The Presence Ontology uses social theory and relational ethnography to model the multiple, idiosyncratic, unbounded interactions of clinical care into a controlled vocabulary that attempts to define the reified elements of *presence* and resultant relationships. While this may appear paradoxical, a relational approach is useful in order to elucidate what would otherwise be a black box for knowledge classification, even if such an approach suggests the limits of ever perfectly transcribing experiential data that is situated, relativistic, and inherently partial. We sought to develop a usable interdisciplinary biomedical ontology that provides a shared language for the subjective components in medicine, without compromising formal rigor and cross-domain interoperability qualities. This approach is resonant with the move toward "social interactionism" in informatics, based on Kaplan's model of communication, control, care, and context[52], and in the pluralistic methods of socio-informatics[41].

Using relational approaches, we modeled abstract entities (e.g. individuals, encounters, characteristics, emotions, qualities, tools) within which domain-specific categories (such as doctors, patients, stethoscopes, waiting rooms, medications, or diseases) can reside. Encounters are interpersonal experiences where concepts like empathy and trust cannot be attributed to one entity but are the result of complex interactions. To this end, we developed the class entity "Relational Quality" that defines qualities emerging from encounters such as empathy, compassion, or trust.

Good patient care is found not on a computer screen but in being truly present with patients[53]. The social sciences theory and ethnographic methods used to develop our ontology achieve a broader reach and greater usability towards developing "good patient care" than existing frameworks discussed in **Section 2**. Our ontology does not solely focus on providers; it suggests that many interactions in healthcare meaningfully modify patient-provider interactions. These warrant further exploration. By modeling both providers and patients as individuals with specific characteristics, the model can help elucidate if providers' individual characteristics (gender, race, age, personality, etc.) and congruence with patients' identities can influence relational experiences like empathy or outcomes like patient satisfaction or clinician burnout[54].

Our study uses relational ethnography to refine the ontology engineering process in order to bring greater specificity and accuracy to our model of subjective healthcare experiences. Recent research in bioinformatics has favored automated observational methods to study the clinical environment. For example, one study suggests incorporating automated techniques like simulation models or tracking sensors to tag healthcare providers and to visualize and analyze clinical workflow in an emergency room[55]. However, automated methods alone miss key features of interactions that render meaning, especially those that are not visible or tangible (such as feelings, qualities, or cognitive models). Space is not the only variable for interactional content. Serious clinical encounters can involve social talk, humor, or self-disclosure—thus appearing less "professional"—while casual hallway conversations can highlight qualities of professionalism that may belie the label of "casual." Thus, using only narrow ethnographic or spatial data cannot provide the breadth that a relational approach allows.

We acknowledge the trade-offs of using a relational approach for ontology engineering: since we seek a wider framework encompassing all possible modifiers of *presence* in clinical interactions, some of our categories are not modeled with enough specificity to provide depth into a specific subcomponent. However, this attempt is the first of its kind. Future work should combine our conceptual breadth with greater depth.

## 5.1 Study Limitations

Although observations of physician-patient interactions informed the ontology, we did not interview patients or elicit their feedback about the ontology. Future work should engage this important community of stakeholders in ontology validation and applications. Similarly, while we envision that our ontology will be applicable to a range of healthcare providers (such as nurses, social workers, respiratory therapists, etc.), the providers on our research team are currently limited to physicians, and physicians author the majority of the domain literature. In future work, we hope to collaborate with a broader array of providers. Finally, despite the diversity of backgrounds on our team, the ontology may not capture all aspects of *presence* related to health equity. We plan a follow-up study to evaluate the Presence Ontology with a racial justice lens, given growing evidence of structural racism and inequitable care in healthcare systems and informatics[56].

## 5.2 Future Directions

We envision the Presence Ontology will be useful to a broad range of end-users, including providers, patients, their families and friends, as well as those who manage, design, or work within healthcare systems or in non-medical fields where connection is analogously important to professional roles and identity. Interactions in healthcare matter, not only for the subjective experiences of patients and providers but also for measurable outcomes such as minimization of medical errors, increased efficiency, equity, and population health. Our ontology can support research on connection in medicine that seeks to make claims about how *presence* affects outcomes such as these.

Future work should involve gathering more information about *presence* in the real world in order to refine the Presence Ontology and adapt it to develop research and/or clinical tools aimed at improving healthcare:

1. *Ethnographic Data Mining:* Patient-provider videotaped encounters or transcripts of audiotaped encounters could generate useful and novel data and meta-data about *presence* in healthcare. The Presence Ontology could be used to annotate transcripts or create a codebook for real-time ethnographic analysis that could be analyzed using machine learning methods, for example.

2. *Documentation of Presence:* The ontology could offer precision and specificity to scene analysis methods using ambient intelligence (combining artificial intelligence and contactless sensors) to assess metaphorical "dark spaces" in medicine and explore the interplay between environment and health behaviors[57].

3. *Clinical Instruments for Presence*: While numerous clinical instruments exist for rating aspects of *presence* such as empathy, burnout, or patient satisfaction, a unified clinical instrument could improve the uptake and measure the success of high-yield, teachable behaviors to improve connection (such as the Presence Center's five recommendations to enhance *presence*[9]). Such an instrument could be used to educate trainees in the art of connection and adapted into a checklist to empower patients about aspects of *presence* that they should expect in encounters.

# 6 CONCLUSION

We have demonstrated a novel classification of the subjective domain of human experience using an ethnographic approach to ontology engineering. Our Presence Ontology synthesizes multiple forms of data and uses relational ethnography to model connection at a high level of abstraction and with clarity. The Presence Ontology focuses on the interpersonal dynamics among providers, patients, and families and friends; the factors that may influence these interactions; and the outcomes they generate. As a result, our conceptual model may have broader reach and greater usability than existing frameworks. A model for *presence* and future applications of our ontology can offer shared agendas and support novel informatics applications to improve human connection in healthcare.

## AUTHOR CONTRIBUTIONS

A.M. conducted the pilot ethnographic study to observe clinical encounters, conducted the broad literature survey, and decided on concepts and relations in the Presence Ontology. M.K. formalized the Presence Ontology in OWL using the Protégé Ontology Editor and published it on BioPortal. Along with A.M., M.M. developed the ontology using anthropological methods. M.M. and A.V. supervised A.M. in iterative ontology development. M.H. led the systematic review of interpersonal interventions geared toward improving presence. C.B. led the ethnographic study of 27 patient-physician encounters and contributed toward development of upper-level concepts in the ontology. R.S. led the qualitative study of trans-disciplinary professional insights about interpersonal connection. D.Z. led the Stanford Presence 5 research team. A.V. and S.T provided insight to refine ontology concepts. A.M. and M.K. prepared the initial manuscript draft. All authors were involved in critical revisions, editing, and manuscript approval.

## ACKNOWLEDGMENTS

The authors acknowledge Michelle B. Bass, PhD, MSI, Population Research Librarian, Stanford School of Medicine, for her coordination of the literature review. Stanford University's Institutional Review Board approved human subjects research led by Maitra (Protocol No. 30711) and led by Brown-Johnson (Protocol No. 42397). We thank Dr. Sheila Lahijani for her helpful input.

## DATA AVAILABILITY

The Presence Ontology is made available via the BioPortal repository of biomedical ontologies at `http://bioportal.bioontology.org/ontologies/PREO`.

## COMPETING INTERESTS

We have no disclosures to report.

## FUNDING

This work was supported by the Arthur Vining Davis Foundation, a grant from the Gordon & Betty Moore Foundation (#6382, Zulman & Verghese, PIs), and by grant R01 GM121724 from the U.S. National Institute of General Medical Sciences.

## REFERENCES

1. Verghese A, Brady E, Kapur CC, Horwitz RI. The Bedside Evaluation: Ritual and Reason. *Ann Intern Med*. 2011;155(8):550-553. doi:10.7326/0003-4819-155-8-201110180-00013
2. Verghese A. Culture shock - Patient as icon, icon as patient. *N Engl J Med*. 2008;359(26):2748-2751. doi:10.1056/NEJMp0807461
3. Kleinman A. Caregiving: the odyssey of becoming more human. *Lancet*. 2009;373(9660):292-293. doi:10.1016/S0140-6736(09)60087-8
4. Kerasidou A, Horn R. Making space for empathy: Supporting doctors in the emotional labour of clinical care Ethics in Clinical Practice. *BMC Med Ethics*. 2016;17(1):8. doi:10.1186/s12910-016-0091-7
5. Smith CK, Polis E, Hadac RR. Characteristics of the initial medical interview associated with patient satisfaction and understanding. *J Fam Pract*. 1981;12(2):283-288.
6. Verghese A, Shah NH, Harrington RA. What this computer needs is a physician humanism and artificial intelligence. *JAMA - J Am Med Assoc*. 2018;319(1):19-20. doi:10.1001/jama.2017.19198
7. Martin SA, Sinsky CA. The map is not the territory: medical records and 21st century practice. *Lancet*. 2016;388(10055):2053-2056. doi:10.1016/S0140-6736(16)00338-X

8. Brown-Johnson C, Schwartz R, Maitra A, et al. What is clinician presence? A qualitative interview study comparing physician and non-physician insights about practices of human connection. *BMJ Open*. 2019;9(11):e030831. doi:10.1136/bmjopen-2019-030831
9. Zulman DM, Haverfield MC, Shaw JG, et al. Practices to Foster Physician Presence and Connection with Patients in the Clinical Encounter. *JAMA - J Am Med Assoc*. 2020;323(1):70-81. doi:10.1001/jama.2019.19003
10. Wood JT. Interpersonal Communication: Everyday Encounters, 8th Edition - 9781285445830 - Cengage.
11. Hersh WR. Medical InformaticsImproving Health Care Through Information. *JAMA*. 2002;288(16):1955-1958. doi:10.1001/jama.288.16.1955
12. Waitzkin H, Britt T. A critical theory of medical discourse: How patients and health professionals deal with social problems. *Int J Heal Serv*. 1989;19(4):577-597. doi:10.2190/L84U-N4MQ-9YAC-D4PP
13. Presson AP, Zhang C, Abtahi AM, Kean J, Hung M, Tyser AR. Psychometric properties of the Press Ganey® Outpatient Medical Practice Survey. *Health Qual Life Outcomes*. 2017;15(1):1-7. doi:10.1186/s12955-017-0610-3
14. Gardner RL, Cooper E, Haskell J, et al. Physician stress and burnout: the impact of health information technology. *J Am Med Informatics Assoc*. 2019;26(2):106-114. doi:10.1093/jamia/ocy145
15. Bakken S. Can informatics innovation help mitigate clinician burnout? *J Am Med Informatics Assoc*. 2019;26(2):93-94. doi:10.1093/jamia/ocy186
16. Israni ST, Verghese A. Humanizing Artificial Intelligence. *JAMA - J Am Med Assoc*. 2019;321(1):29-30. doi:10.1001/jama.2018.19398
17. Kohane IS. Bioinformatics and Clinical Informatics: The Imperative to Collaborate. *J Am Med Informatics Assoc*. 2000;7(5):512-516. doi:10.1136/jamia.2000.0070512
18. Donnelly K. {SNOMED-CT}: The advanced terminology and coding system for eHealth. *Stud Health Technol Inform*. 2006;121:279-290.
19. Unertl KM, Schaefbauer CL, Campbell TR, et al. Integrating community-based participatory research and informatics approaches to improve the engagement and health of underserved populations. *J Am Med Informatics Assoc*. 2015;23(1):60-73. doi:10.1093/jamia/ocv094
20. Unertl KM, Novak LL, Johnson KB, Lorenzi NM. Traversing the many paths of workflow research: Developing a conceptual framework of workflow terminology through a systematic literature review. *J Am Med Informatics Assoc*. 2010;17(3):265-273. doi:10.1136/jamia.2010.004333
21. Forsythe DE, Buchanan BG. Knowledge Acquisition for Expert Systems: Some Pitfalls and Suggestions. *IEEE Trans Syst Man Cybern*. 1989;19(3):435-442. doi:10.1109/21.31050
22. Forsythe DE. Engineering Knowledge: The Construction of Knowledge in Artificial Intelligence. *Soc Stud Sci*. 1993;23(3):445-477. doi:10.1177/0306312793023003002
23. Belkin NJ, Brooks HM, Daniels PJ. Knowledge elicitation using discourse analysis. *Int J Man Mach Stud*. 1987;27(2):127-144. doi:10.1016/S0020-7373(87)80047-0
24. Ventres WB, Frankel RM. Shared presence in physician-patient communication: A graphic representation. *Fam Syst Heal*. 2015;33(3):270-279. doi:10.1037/fsh0000123
25. Larson EB, Yao X. Clinical empathy as emotional labor in the patient-physician relationship. *J Am Med Assoc*. 2005;293(9):1100-1106. doi:10.1001/jama.293.9.1100
26. Hochschild AR. *The Commercialization of Intimate Life: Notes from Home and Work*. Berkeley: Univ of California Press, 2003; 2003.
27. PubMed. https://pubmed.ncbi.nlm.nih.gov/.
28. Frank JD, Frank JB. *Persuasion and Healing: A Comparative Study of Psychotherapy*. JHU Press; 1993.
29. Ainsworth-Vaughn N. *Claiming Power in Doctor-Patient Talk*. Oxford University Press on Demand; 1998.
30. Foucault M. *The Birth of the Clinic*. Routledge; 2012.
31. Jain SL. *Malignant: How Cancer Becomes Us*. Univ of California Press; 2013.
32. Mol A. *The Logic of Care: Health and the Problem of Patient Choice*. Routledge; 2008.
33. KLEINMAN A, EISENBERG L, GOOD B. Culture, Illness, and Care. *Ann Intern Med*. 1978;88(2):251-258. doi:10.7326/0003-4819-88-2-251

34. Forsythe D. *Studying Those Who Study Us: An Anthropologist in the World of Artificial Intelligence*. Stanford University Press; 2001.
35. Ozdalga E, Ozdalga A, Ahuja N. The Smartphone in Medicine: A Review of Current and Potential Use Among Physicians and Students. *J Med Internet Res*. 2012;14(5):e128. doi:10.2196/jmir.1994
36. Haverfield MC, Tierney A, Schwartz R, et al. Can Patient–Provider Interpersonal Interventions Achieve the Quadruple Aim of Healthcare? A Systematic Review. *J Gen Intern Med*. 2020;35(7):2107-2117. doi:10.1007/s11606-019-05525-2
37. Embase. https://www.embase.com.
38. PsycINFO. https://www.apa.org/pubs/databases/psycinfo.
39. Sikka R, Morath JM, Leape L. The quadruple aim: Care, health, cost and meaning in work. *BMJ Qual Saf*. 2015;24(10):608-610. doi:10.1136/bmjqs-2015-004160
40. Desmond M. Relational ethnography. *Theory Soc*. 2014;43(5):547-579. doi:10.1007/s11186-014-9232-5
41. Wulf V, Pipek V, Randall D, Rohde M, Schmidt K, Stevens G. *Socio-Informatics*. Oxford University Press; 2018.
42. Handwerker PW. *Quick Ethnography: A Guide to Rapid Multi-Method Research*. Rowman Altamira; 2001.
43. Schwartz R, Haverfield MC, Brown-Johnson C, et al. Transdisciplinary Strategies for Physician Wellness: Qualitative Insights from Diverse Fields. *J Gen Intern Med*. 2019;34(7):1251-1257. doi:10.1007/s11606-019-04913-y
44. Tudorache T, Noy NF, Tu S, Musen MA. Supporting Collaborative Ontology Development in Protégé. In: Springer, Berlin, Heidelberg; 2008:17-32. doi:10.1007/978-3-540-88564-1_2
45. Kamdar MR, Tudorache T, Musen MA. A systematic analysis of term reuse and term overlap across biomedical ontologies. *Semant Web*. 2017;8(6):853-871. doi:10.3233/SW-160238
46. Hastings J, Ceusters W, Smith B, Mulligan K. The emotion ontology: Enabling interdisciplinary research in the affective sciences. In: *Lecture Notes in Computer Science (Including Subseries Lecture Notes in Artificial Intelligence and Lecture Notes in Bioinformatics)*. Vol 6967 LNAI. Springer, Berlin, Heidelberg; 2011:119-123. doi:10.1007/978-3-642-24279-3_14
47. Whetzel PL, Noy NF, Shah NH, et al. BioPortal: Enhanced functionality via new Web services from the National Center for Biomedical Ontology to access and use ontologies in software applications. *Nucleic Acids Res*. 2011;39(SUPPL. 2):W541-W545. doi:10.1093/nar/gkr469
48. Jonquet C, Shah NH, Youn CH, Callendar C, Storey M-A, Musen MA. NCBO Annotator: Semantic Annotation of Biomedical Data. In: *Proceedings of the International Semantic Web Conference (ISWC), Poster and Demo Session, Vol. 110.* ; 2009.
49. Sinclair S, McClement S, Raffin-Bouchal S, et al. Compassion in Health Care: An Empirical Model. *J Pain Symptom Manage*. 2016;51(2):193-203. doi:10.1016/j.jpainsymman.2015.10.009
50. Anderson RT, Camacho FT, Balkrishnan R. Willing to wait? The influence of patient wait time on satisfaction with primary care. *BMC Health Serv Res*. 2007;7(1):1-5. doi:10.1186/1472-6963-7-31
51. Smith B, Ceusters W. Ontological realism: A methodology for coordinated evolution of scientific ontologies. *Appl Ontol*. 2010;5(3-4):139-188. doi:10.3233/AO-2010-0079
52. Kaplan B. Evaluating informatics applications—some alternative approaches: theory, social interactionism, and call for methodological pluralism. *Int J Med Inform*. 2001;64(1):39-56. doi:https://doi.org/10.1016/S1386-5056(01)00184-8
53. Verghese A. The importance of being. *Health Aff*. 2016;35(10):1924-1927. doi:10.1377/HLTHAFF.2016.0837
54. Gleichgerrcht E, Decety J. Empathy in Clinical Practice: How Individual Dispositions, Gender, and Experience Moderate Empathic Concern, Burnout, and Emotional Distress in Physicians. Zalla T, ed. *PLoS One*. 2013;8(4):e61526. doi:10.1371/journal.pone.0061526
55. Vankipuram A, Traub S, Patel VL. A method for the analysis and visualization of clinical workflow in dynamic environments. *J Biomed Inform*. 2018;79:20-31. doi:10.1016/j.jbi.2018.01.007
56. Cahan EM, Hernandez-Boussard T, Thadaney-Israni S, Rubin DL. Putting the data before the algorithm in big data addressing personalized healthcare. *npj Digit Med*. 2019;2(1):78. doi:10.1038/s41746-019-0157-2
57. Haque A, Milstein A, Fei-Fei L. Illuminating the dark spaces of healthcare with ambient intelligence. *Nature*. 2020;585(7824):193-202. doi:10.1038/s41586-020-2669-y